A systematic approach to identify and evaluate

missing data patterns and mechanisms in

multivariate educational, social, and behavioral research

Adam Davey[1][¶][*] and Ting Dai[2][¶]

[1] Department of Behavioral Health and Nutrition, University of Delaware, Newark, Delaware,

United States of American

[2] Department of Educational Psychology, University of Illinois at Chicago, Chicago, Illinois,

United States of America

* Corresponding author

E-mail: davey@udel.edu (AD)

[¶]These authors contributed equally to this work.

# Abstract

Methods for addressing missing data have become much more accessible to applied researchers. However, little guidance exists to help researchers systematically identify plausible missing data mechanisms in order to ensure that these methods are appropriately applied. Two considerations motivate the present study. First, psychological research is typically characterized by a large number of potential response variables that may be observed across multiple waves of data collection. This situation makes it more challenging to identify plausible missing data mechanisms than is the case in other fields such as biostatistics where a small number of dependent variables is typically of primary interest and the main predictor of interest is statistically independent of other covariates. Second, there is growing recognition of the importance of systematic approaches to sensitivity analyses for treatment of missing data in psychological science. We develop and apply a systematic approach for reducing a large number of observed patterns and demonstrate how these can be used to explore potential missing data mechanisms within multivariate contexts. A large scale simulation study is used to guide suggestions for which approaches are likely to be most accurate as a function of sample size, number of factors, number of indicators per factor, and proportion of missing data. Three applications of this approach to data examples suggest that the method appears useful in practice.

# Introduction

Missing (or "incomplete" or "nonresponse") data are ubiquitous in the educational, social, and behavioral research settings (1). In recent years, methods for handling missing data have become increasingly widely available to non-methodologists, and many software packages include routines for direct maximum likelihood or multiple imputation approaches that previously required considerable methodological sophistication to implement correctly. Systematic exploration of missing data mechanisms is critical to selecting and applying the most appropriate statistical techniques for analysis of incomplete data (2,3).

Within the context of randomized controlled trials, which feature an experimental design, one or a very small number of discrete outcome variables and a well-defined study end-point, reasonable advice and guidance is available, e.g., (3–8). Recommendations tend to focus around investigating a series of sensitivity analyses in order to better understand how assumptions about the missing data mechanism might affect study results (7).

It is important for researchers to understand how the results obtained appear to depend on the implicit or explicit assumptions being made about the unobserved values. In both experimental and observational contexts, there is also an emerging emphasis on the role of sensitivity analyses in the analysis of missing data (4,8–10). There are also sophisticated examples of the same approach in observational studies with a single response variable of interest, e.g., (11).

However, in educational, social and behavioral contexts, which are most often observational and multivariate, very little systematic and concrete guidance currently exists. Researchers looking for guidance in identifying patterns and predictors of nonresponse will typically find few examples. One exception to this is (12), which first identified discrete patterns of missing data and then examined variables associated with the probability of nonresponse in order to adjust models for dropout. However, the method in that paper was not specified in sufficient detail to provide a broadly useful and replicable framework for analysis of incomplete data, nor was the performance of alternative approaches considered and evaluated in the context of simulated data where the underlying true structure was known.

Two notions of missing data are important -- *missing data patterns* and *missing data mechanisms*. Given $k$ different variables, a missing data pattern is represented by a vector of length $k$ indicating observed/missing status on each variable. A study with $k$ different variables has the potential to result in up to $2^k - 1$ different meaningful patterns of missing data. The formula subtracts 1, because observations missing on all $k$ variables contribute no information. On the other hand, the pattern, "observed on all $k$ variables," is included in the overall consideration of missing data in a study and represents the subset of observations remaining after listwise deletion.

Missing data mechanisms are most often considered in terms of Rubin's (1976) taxonomy, which describes missing data as being Missing completely at random (MCAR), Missing at random (MAR), or Missing not at random (MNAR). Data are said to be MCAR if the probability of missingness does not depend on the values of covariates ($X$), auxiliary ($V$), or outcome ($Y$) variables ($X, V, Y$), i.e., $Pr(M|X, V_{Obs}, V_{Mis}, Y_{Obs}, Y_{Mis}) = Pr(M)$. Data are typically only MCAR when missing data result from *a priori* design decisions such as planned missing designs, e.,g., XABC design (13), matrix sampling (14). In contrast, data are MAR when missingness is independent of missing responses $Y_{Mis}$ and, conditioned on observed responses $Y_{Obs}, V_{Obs}$ and covariates $X$, i.e., $Pr(M|X, V_{Obs}, V_{Mis}, Y_{Obs}, Y_{Mis}) = Pr(M|X, V_{Obs}, Y_{Obs})$. Both MCAR and MAR data are said to be *ignorable*, because it does not need to be modeled in order to draw inferences. When missingness or nonresponse depends on the values of missing variables, conditioning on the observed variables, data are said to be MNAR, i.e., $Pr(M|Y_{Obs}, Y_{Mis}) = Pr(M|Y_{Mis})$. MNAR data are said to be *nonignorable* and must be modeled explicitly.

The present article develops a framework for beginning with first reducing *missing data patterns* as a way to guide systematic consideration of *missing data mechanisms* in situations where the number of missing data patterns is potentially very large.

## Data Reduction

Several methods for data reduction are familiar to scientists, including principal component analysis and factor analysis (or principal axis factoring). Principal component analysis (PCA) seeks to reduce the dimensionality of a data set containing a large number of correlated variables in a way that retains much of the variation present in the full data set. This is accomplished by transforming to a new set of uncorrelated variables or principal components. Principal components are extracted such that the first few retain most of the variation present in the original set of variables. In other words, weights are selected to form linear combinations of the original variables in order to construct new orthogonal variables which capture a maximal variance in the original variables for the reduced number of composites. Thus, in PCA the emphasis is on capturing variance rather than covariance or explaining relations between constructs. (15)

In contrast, factor analysis (FA) hypothesizes an underlying set of "common factors" that can explain, or account for, the associations among observed variables. (16) Factor analysis attempts to achieve a reduction in dimensions by invoking a model relating observed variables to a smaller number of hypothetical or latent variables. In PCA, the goal is to find optimal ways of creating subsets of variables without imposing a model or structure on the data. In contrast, the goal of FA is to identify the underlying structure with the goal of capturing covariance rather than variance of the variables.

In this paper, we emphasize a PCA approach, because the approach is deliberately empirical rather than theoretical and delivers a mathematically unique solution. However, the methods we apply can also be used with factor analysis with suitable modifications when researchers believe that is the most appropriate method (i.e., when there is interest in identifying and understanding underlying sources of missing data rather than simply ensuring that they are appropriately modeled in an analysis). While it is possible that a researcher may know (or suspect) in advance a small number of common causes for missing data in their study, more typically this is something to be determined from the data themselves. For example, what is the smallest number of missing data patterns that can account for a given proportion of variance in missing data patterns?

## Determining the Number of Components

There is a vast literature on criteria for selecting an appropriate number of principal components (15). One of the most widely applied is Kaiser's criterion (17), which retains all components with eigenvalues greater than one (i.e., components explaining at least as much variance as a typical standardized item). Kaiser's method is the default approach in most statistical packages (e.g., SPSS), although some studies have suggested this approach may lead to over-extraction of components in many applications, cf. (18).

Kaiser's method ignores the fact that eigenvalues are sorted from largest to smallest, thus capitalizing on chance differences in sampling variance. To address this issue, parallel analysis (19) uses data simulated under an independence assumption to subtract out this sampling error variance, and solutions can be evaluated using several different criteria (e.g., mean, median, 95%ile).  A wide variety of simulation studies (18) has suggested that parallel analysis provides an unbiased estimate of the number of underlying components and works well in practice.

Quite recently, parallel analysis has been extended to consider data generated under data structures with varying numbers of components in the Comparison Data (CD) method (18), with additional components being added to the simulated data for as long as they produce better agreement with the structure of the original data. We do not consider this method here because it: a) can be quite computationally-intensive and b) generally agrees quite well with results from parallel analysis; however, it represents a potentially important addition to the methods available for determining the number of components to be retained.

In empirical contexts, many researchers rely on scree plots in order to determine the number of components to retain in an analysis. A plot of eigenvalues against the eigenvalue number is used to identify an "elbow" or "large gap" in the data at the point where the useful "signal" degenerates into noise, or "scree."  However, this method provides no definite quantitative cutoffs, and hence is difficult to use for empirical evaluation.  The method of profile likelihood recently presented by Zhu and Ghodsi (20) attempts to address this shortcoming by quantifying the number of retained components that maximizes the observed data likelihood, thus providing an empirical (and automatic) method for determining the point at which a "gap" or "elbow" occurs within a scree plot.

Finally, an empirical Kaiser criterion (EKC) has recently been presented in the literature (21). This approach, which is based on statistical theory and accounts for the serial nature of eigenvalues. In a Monte Carlo study, the EKC approach generally performed at least as well as parallel analysis, particularly with larger sample sizes and a smaller number of variables.

For the current application, however, we require evaluation of methods for component/factor structure with binary, rather than continuous, indicators (i.e., missing/observed). A recent simulation study by Finch (22) indicated that, while parallel analysis generally performed well, it was not as effective in factor analysis with dichotomous indicators. Finch recommended a confirmatory factor analysis approach using the root mean square error of approximation to determine the number of factors. However, the results also suggested that lack of convergence may often be expected under the kinds of circumstances faced by applied researchers (e.g., smaller sample sizes and larger number of variables). Indeed, the confirmatory approach is not even identified unless each factor is represented by at least three manifest indicators, a situation that may not be realistic in terms of missing data patterns. For this reason, we do not consider this approach further.

## Materials and methods

This paper proceeds in the following way. Study 1 reports results of a large scale simulation study designed to reliably determine the underlying dimensionality of a set of binary indicators. In Study 2, building on this information, we develop and evaluate an eight step process for exploring the missing data mechanism and apply it to three distinct data sets where the underlying missing data mechanism is unknown or only partially known.

# Study 1: A Simulation Study

## Purpose

The purpose of Study 1 was to evaluate the performance of different methods for determining the correct number of components/factors underlying a set of missing data patterns under a range of conditions typical for multivariate social, behavioral, and educational research.

## Design

In order to consider a range of conditions typical of many real-world missing data applications, e.g., (1,23), we used a factorial design. Between-subjects factors were: (1) the number of components (1, 3, 5, 10), (2) number of items per component (3, 5, 10), (3) sample size (100, 250, 1000), (4) probability of missing values (10%, 25%, 50%). Within-subjects factors were: (5) type of correlation matrix analyzed (Pearson's $r$ or tetrachoric $\rho$), and (6) analysis type (PCA or PAF).

## Outcomes

We evaluated criteria for selecting the number of components to retain. These included: 1) empirical Kaiser's criterion, 2) Kaiser's criterion, 3) parallel analysis with 95%ile criterion, and 4) profile likelihood.We consider a method successful in a specific condition if it recovers the correct number of components in at least 95% of replications.

## Procedure

We constructed 1000 replications in each condition. Population correlations were set at .7 within variables on the same components and .3 between variables on different components, and all data were drawn from multivariate normal distributions. Variables were dichotomized based on whether observed values exceeded the threshold associated with the probability of missing data for that condition.

For each replication, raw data were converted to correlation matrices (Pearson's $r$ and tetrachoric $\rho$) and analyzed by principal component analysis (PCA) and principal axis factoring (FA). The number of components/factors indicated by each criterion was calculated for each replication. Dichotomous indicators were then constructed to indicate, for each replication and criterion, whether the correct number of factors/components was recovered. Ordering from overall highest proportion correctly recovered to overall lowest proportion correctly recovered (some criteria performed identically in some conditions), we constructed a nominal indicator of which criterion performed best (1 = parallel analysis, 2 = empirical Kaiser, 3 = Kaiser, and 4 = profile likelihood).

## Statistical Analysis

Based on descriptive analyses, we determined that the correct number of factors/components were correctly extracted more often via principal component analysis and Pearson correlation. Thus only these results are presented in the body of the text, with data for the remaining conditions available via online supplementary materials. Factorial Analysis of variance (ANOVA) was used to investigate association of each factor with recovery of the correct number of factors/components, in preference to logistic regression owing to the large number of conditions where some criteria performed perfectly, which can lead to estimation difficulties with likelihood-based techniques. We followed up these analyses using cross-validated recursive partitioning (24) in order to guide researchers to the most effective criterion for a given condition.

## Software

The R statistical package (25) was used for all analyses. All code and data generated are available at: https://github.com/soy2padre/misspatt.

# Results

Initial inspection of the results suggested that PCA with Pearson's correlation matrices generally performed at least as well as other approaches while also providing the highest convergence rates and so we present results only for this condition here. (Full results are available in supplementary materials.) Because of the large number of replications, ANOVAs indicated that all model effects were statistically significant at p<.05, including the 5-way interactions. For this reason, we summarize the key findings and influences for each criterion.

Figs 1-9 present the proportion of replications in which each method recovered the correct number of components by simulation condition for principal component analysis and Pearson correlation. Overall, parallel analysis most often recovered the correct underlying structure (87.8%), followed by empirical Kaiser criterion (79.3%), Kaiser (73.3%), and profile likelihood (67.4%). However, these aggregate results masked considerable variation across study conditions. Additionally, there were striking differences in the importance of each factor in determining the situations where each criterion was more and less likely to yield correct results.

**Fig. 1. Proportion of replications recovering the correct number of components (N=100, 10% Missing).**

**Fig. 2. Proportion of replications recovering the correct number of components (N=100, 25% Missing).**

**Fig. 3. Proportion of replications recovering the correct number of components (N=100, 50% Missing).**

**Fig. 4. Proportion of replications recovering the correct number of components (N=250, 10% Missing).**

**Fig. 5. Proportion of replications recovering the correct number of components (N=250, 25% Missing).**

**Fig. 6. Proportion of replications recovering the correct number of components (N=250, 50% Missing).**

**Fig. 7. Proportion of replications recovering the correct number of components (N=1000, 10% Missing).**

**Fig. 8. Proportion of replications recovering the correct number of components (N=1000, 25% Missing).**

**Fig. 9. Proportion of replications recovering the correct number of components (N=1000, 50% Missing).**

## Parallel analysis

Recall that our factorial design included the following conditions -- 3 sample sizes, 4 numbers of components, 3 numbers of items per component, 3 probabilities of missing data, which created a total of 108 conditions given PCA and Pearson correlation. Although it performed best of all selection methods considered, performance of parallel analysis was driven primarily by sample size. Parallel analysis correctly covered the number of components at least 95% of the time under all 36 conditions when sample size was 1000. It correctly recovered the number of parameters at least 95% of the time in 31 of 36 conditions when sample size was 250 (missing 4 of 9 conditions when the number of components was 10, and 1 of 9 conditions when the number of components was 5). Parallel analysis recovered the correct number of components at least 95% of the time in only 14 of 36 conditions where sample size was 100. This was also the only sample size for which the approach converged in fewer than 1000 of the replications. With a sample size of 100, the method was least effective in conditions where the proportion of missing data was low (10%), recovering the correct number of components in at least 95% of the replications in only 2 of 12 conditions, 6 of 12 conditions with 25% missing data, and 6 of 12 conditions with 50% missing data.

## Empirical Kaiser

Performance of the empirical Kaiser criterion was highest with larger sample sizes, recovering the correct number of components at least 95% of the time in 34 of 36 conditions when sample size was 1000, missing only the conditions with 10 items per component with 10 components (0% correct) and 5 components (94.8% correct). EKC recovered the correct number of components at least 95% of the time in 22 of 36 conditions when sample size was 250, performing more poorly with a larger number of items per component and, to a lesser extent, a larger number of components. EKC recovered the correct number of components at least 95% of the time in 10 of 36 conditions when sample size was 100, performing more poorly with a larger number of items per component and, to a lesser extent, a larger number of components.

## Kaiser

Performance of the Kaiser criterion followed an essentially identical pattern to that of the empirical Kaiser criterion, but recovered the correct number of components a slightly lower proportion of the time, overall.

## Profile likelihood

Profile likelihood showed a relatively distinct pattern of performance than the other three methods considered for selecting the correct number of factors. Specifically, the number of factors was important with profile likelihood. Profile likelihood recovered the correct number of components at least 95% of the time under all 36 conditions involving a single component. Profile likelihood recovered the correct number of components at least 95% of the time under 12 of 36 conditions involving 3 components. Profile likelihood recovered the correct number of components at least 95% of the time under 6 of 36 conditions involving 5 components. Profile likelihood recovered the correct number of components at least 95% of the time under just 1 of 36 conditions involving 10 components.

## Guidance for selecting a criterion

Although parallel analysis generally outperformed the other methods under most circumstances, we used cross-validated recursive partitioning to develop a decision tree to guide researchers to conditions where an alternative approach might perform better. The tree was pruned using a complexity parameter criterion of 0.02 and is shown in Fig 10 below.

**Fig. 10. Decision tree for selecting a criterion for the number of components.**

The decision tree suggests that parallel analysis should be selected except under conditions where: the number of items per component is 5 or more, the sample size is less than 1000, 3 or more components are to be recovered. Then, if the sample size is 250, the empirical Kaiser criterion should be slightly preferred and if sample size is 100, then the Kaiser criterion should be used. It is noteworthy, however, that when the EKC criterion is recommended the difference in correct determination rates between EKC and regular Kaiser criterion is less than 0.1% in absolute terms (98.3% vs 98.2%).

## Discussion of Study 1

Results from Study 1 strongly suggest that parallel analysis should generally be preferred when reducing a potentially large number of missing data patterns into a smaller subset for consideration in terms of the corresponding missing data mechanism(s). Recursive partitioning suggested that a regular Kaiser criterion will be more effective than parallel analysis under specific conditions where the sample size is less than 1000, the items per component are 5 or more, and 3 or more components are represented.

Having established that a potentially large number of distinct missing data patterns can accurately be reduced to a considerably smaller number of missing data patterns suitable for examining their associated missing data mechanisms, we provide a set of guidance for how to do so in practice with the steps outlined in Table 1 (12). Step 1 is to construct dichotomous indicators of all variables with missing data (1 = missing, 0 = observed). Step 2 is to perform a PCA on missing data indicators obtained at Step 1. Step 3 is to determine the number of components to retain using a method with the best performance (e.g., parallel analysis) based on the design characteristics (e.g., sample size, the number of items per component, and percentage of missing data) by referring to our Study 1 findings. Step 4 is to extract the desired number ($k$) of principal components and calculate predicted component scores. Step 5 is to dichotomize component scores (1 = missing, 0 = observed). The cutoff can be determined empirically, such as by examining the distribution of components scores, or else based on a split at zero.

**Table 1. Recommended Steps for Examining Missing Data Mechanism in a Multivariate Context.**

| Step | Task and Consideration |
|:---:|:---|
| 1 | Create dichotomous indicators of all variables with missing data (1 = missing, 0 = observed) |
| 2 | Run principal component analysis on the dichotomous missing data indicators |
| 3 | Select criteria (e.g., parallel analysis) to determine the number of principal components to retain ($k$) |
| 4 | Extract desired number of components and calculate predicted component scores |
| 5 | Dichotomize principal component scores to indicate main missing patterns (missing = 1, observed = 0) |
| 6 | Examine distributions of observed variables and missing data indicators by each dichotomized component score and known characteristics of study design |
| 7 | Predict dichotomized component scores from observed variables and/or missing data indicators using logistic regression to evaluate MNAR |
| 8 | If no clear pattern emerges, repeat Steps 6-7 by some stratifying variable (e.g., treatment condition, gender) |

By Step 5, the unwieldy number of missing data patterns have been reduced to $k$ main missing patterns as indicated by the principal components. The next steps involve evaluating the obtained reduced missing patterns and understanding the meaning and plausible causes for missing, using all study variables.

Step 6 is to examine -- using *t*-test, chi-square, etc. -- the distributions of observed variables and their corresponding missing data indicators by each component score and each known characteristic of study (e.g., refresher cohort, mortality, school dropout). The results of this preliminary screening show whether or not data differ by a main missing data pattern as indicated by the component score. Doing so by the known characteristics of study design also provides an evaluation of the external validity of the main missing patterns. Step 7 is to predict dichotomized component scores from observed variables or their corresponding missing data indicators using binomial logistic regression or similar methods. The results show the extent to which each main missing data pattern -- as indicated by the component scores -- depends on the observed variables where data are either missing or observed, suggesting the possibility of a NMAR mechanism. If the pattern is still difficult to interpret, in Step 8 stratify the sample by variables (e.g., gender, treatment condition) and then repeat Steps 6 and 7 for each subgroup. We recommend this step in all situations with an experimental manipulation and/or stratification variable. This is especially important for any models that will ultimately include evaluation of one or more interaction terms.

## Study 2: Applications

The value of any method lies in its performance under real-world circumstances. In a second study, we apply the systematic approach described above based on findings from Study 1 to analyze and understand the missing data in three different data sets: 1) example 11.4 from the Mplus User's Guide (26), psychiatric clinical trial data analyzed by Hedeker and Gibbons (27), and data on depressive symptoms in the Health and Retirement Study (HRS; (28)). In these data sets the missing data mechanisms were unknown to us, but partially inferrable. In the remainder of this section, we introduce each data set, report the application of our systematic approach (i.e., the eight steps shown in Table 1) for handling missing data, present the results, and interpret the findings in relations to what is known or can be discerned with regard to missing data in the original study. All analyses of Study 2 were conducted with R (25).

## Application 1: M*plus* Data Example 11.4
### Data

We used the data set, Example 11.4, from the software package, M*plus* Version 8 (26), to test our approach to handling missing data. Example 11.4 in the User's Guide of Mplus is a demonstration of using M*plus* to conduct pattern mixture modeling with MNAR.

To fully describe the missingness of each application data set, we report on the missingness by variable, by sample size, and by missing pattern. The six variables, y0 to y5, of the original data set represent six waves of observed data ("Mplus data set" hereafter). All cases are observed on y0 and so it is excluded from our approach since a missing data for it has no variance. The other five variables have 26.4-38.5% missing data. Of the 1,000 cases in the Mplus data set, 29.9% have no missing data, and 70.1% have at least one pattern of missing data. Given 5 variables with missing values, the maximum number of missing patterns in this data set would be 31 (i.e., $2^5 - 1$). As simulated by the Mplus team, this data set does have 31 different missing data patterns (see Table 2 for missing patterns with the highest and lowest frequencies).

**Table 2. Missing patterns and frequencies with application data sets.**

| Freq. Rank | Missing on Variables | # of Var. with Mis. | Freq. | % of All Cases |
|---|---|---|---|---|
| Mplus ($N$ = 1,000) | $m1 - m5$ | | | |
| 1 | 00000 | 0 | 299 | 29.9% |
| 2 | 00100 | 1 | 74 | 7.4% |
| 3 | 10000 | 1 | 70 | 7.0% |
| 4 | 01000 | 1 | 60 | 6.0% |
| 5 | 00001 | 1 | 39 | 3.9% |
| 28 | 11010 | 3 | 7 | 0.7% |
| 29 | 10010 | 2 | 6 | 0.6% |
| 30 | 10010 | 2 | 6 | 0.6% |
| 31 | 11001 | 3 | 6 | 0.6% |
| 32 | 10011 | 3 | 5 | 0.5% |
| H&G ($N$ = 437) | IMPS79.0m $-$ IMPS79.6m | | | |

| 1 | 0000 | 0 | 312 | 71.4% |
|---|------|---|-----|-------|
| 2 | 0001 | 1 | 53 | 12.1% |
| 3 | 0011 | 2 | 45 | 10.3% |
| 4 | 0010 | 1 | 13 | 3.0% |
| 5 | 0100 | 1 | 5 | 1.1% |
| 6 | 1000 | 1 | 3 | 0.7% |
| 7 | 0111 | 3 | 3 | 0.7% |
| 8 | 0110 | 2 | 2 | 0.5% |
| 9 | 0101 | 2 | 1 | 0.2% |

| CES-D ($N$ = 27,301) | DAm1 SOm1 PAm1 – DAm10 SOm10 PAm10 | | | |
|---|---|---|---|---|
| 1 | 000000000000000000000000000000 | 0 | 4971 | 18.2% |
| 2 | 111111111000000000000000000000 | 9 | 3310 | 12.1% |
| 3 | 111111111111111111000000000000 | 18 | 2340 | 8.6% |
| 4 | 111111111000111111111111111111 | 27 | 1274 | 4.7% |
| 5 | 111111111000000111111111111111 | 24 | 1069 | 3.9% |

| 6 | 00011111111111111111111111111111 | 27 | 928 | 3.4% |
| 894 | 1111111111111.0011111111111111111 | 28 | 1 | 0.004% |
| 895 | 11111111111111110011111111111111 | 28 | 1 | 0.004% |
| 896 | 11111111111111111111111111111001 | 28 | 1 | 0.004% |
| 897 | 11111111110111111111111111111111 | 29 | 1 | 0.004% |
| 898 | 11111111110111111111111111111111 | 29 | 1 | 0.004% |

*Note.* Freq =Frequency. Mis = Missing Data. "1" indicates missing, "0" indicates observed.

The characteristics of the Mplus data set (e.g., relatively large sample size, $N = 1000$, a small number of variables, and complex missing data patterns) make it an interesting starting point for testing our approach to handling and understanding missing data that were not generated as part of our simulation study, because several methods can be expected to perform well under these circumstances, providing opportunities to achieve consensus. Additionally, because they are also simulated data, it is likely that the underlying structure can be inferred with considerable accuracy even though it is unknown in advance.

## Procedure

We sought to reduce the missing data patterns from 31 to a smaller number and consider our recommended approach. First, we recoded y1 to y5 into five corresponding missing data indicators, m1 to m5 (Missing = 1, Observed = 0), performed a principal component analysis on the missing data indicators, used the four methods to determine the number of principal components to extract, and decided the most appropriate number of components to extract from these data (Steps 1-3). Next, we extracted the component, calculated the predicted component scores, and dichotomized the principal component scores to make a binary variable that indicates data missing or not (Steps 4-5). To examine how well this missing pattern captures the overall missingness on y1 to y5, we carried out Steps 6-8 of our systematic approach using the observed missing data indicators m1-m5.

## Results

In Study 1, all four determination criteria -- EKC, Kaiser, parallel analysis, and profile likelihood -- correctly determined the number of components for 100% of the simulated data by the same characteristics as the Mplus data set -- $N = 1000$, 1 principal component, 5 variables per component (see Supplementary Table 1). Thus, we expected 1 principal component to be extracted with the Mplus data set (i.e., $N = 1000$, 5 variables, and 26.4–38.4% of missing) regardless of criterion selected. We conducted principal component analysis on the five missing data indicators, m1 to m5, and found that the four criteria unanimously indicated 1 component (Table 2). As presented in Table 3, all indicators loaded strongly on the extracted component; standardized loadings are all above .42 and four indicators loaded on the component by .80 or higher. A total of 429 (42.9% of all cases) have missing data on this pattern.  This finding supports our expectations based on Study 1, and provides empirical support for the Study 1 findings.

**Table 3. Number of components determined by 4 methods for each application data set.**

| Application Data Set | EKC | Kaiser | Parallel Analysis | Profile Likelihood |
|---|---|---|---|---|
| M*plus* Example 11.4 | **1** | **1** | **1** | **1** |
| Hedeker & Gibbons (1997) | **1** | **1** | **1** | 1 |
| CES-D Data of the HRS | **4** | **4** | **4** | 2 |

*Note*. Boldface indicates in Study 1 a method correctly determines 100% of simulated data with similar characteristics to the corresponding application data set.

We saved the estimated principal component scores, and recoded the component scores into a dichotomous variable, *Component 1* (below or equal to 0 = 0, Not Missing; above 0 = 1, Missing). To examine how well component 1 describes the missingness on y1-y5, we conducted logistic regressions of component 1 on the observed missing indicators, m1 to m5. We estimated 5 simple logistic regressions of component 1 on m1, m2, m3, m4 and m5, individually, and then a $6^{th}$ multiple regression of component 1 on all five missing indicators. We found that although all were significant predictors (Table 4), none of the missing indicators was able to completely predict component 1 as a sole predictor (i.e., correct classification rate below 85%; Table 5). However, in Model 6, m1 to m5 together correctly predicted component 1with 100% accuracy, suggesting that the reduced missing pattern completely describes the overall missingness of the M$plus$ data set, i.e., $logit(Y_{Mis}) = -240 + 110 \times m1 + 156 \times m2 + 176 \times m3 + 156 \times m4 + 165 \times m5$. This offers support for the idea that a single pattern underlies missing data on all five observed variables, and that our blinded analysis of the missing data mechanism derived from 31 distinct missing data patterns allows us to model the probability of nonresponse with 100% accuracy.

**Table 4. Application 1: Mplus data set, steps 2-3: Missing data indicator component loadings.**

| Component Indicator | Component 1 |
|---|---|
| m1 | .429 [c] |
| m2 | **.559 [c]** |
| m3 | **.679 [c]** |
| m4 | **.707 [c]** |
| m5 | **.724 [c]** |

[a] $p < .05$, [b] $p < .01$, [c] $p < .001$.  Boldface indicates a loading > .550.

**Table 5. Application 1: Mplus data set, Steps 6-8: Logistic regressions of the component on missing data indicators.**

| Model | Parameter | Coefficient Estimate | S.E. | Log Likelihood | LR $x^2$ | Pseudo $R^2$ | Area under ROC | Sensitivity (%) | Specificity (%) | Correctly Classified (%) |
|---|---|---|---|---|---|---|---|---|---|---|
| 1 | m1 | 1.776 | 0.160 | -613.675 | 138.71 | .1015 | .6648 | 45.22 | 87.74 | 69.50 |
| | Intercept | -0.757 | 0.079 | | | | | | | |
| 2 | m2 | 2.447 | 0.168 | -550.682 | 264.70 | .1938 | .7353 | 57.58 | 89.49 | 75.80 |
| | Intercept | -1.032 | 0.086 | | | | | | | |
| 3 | m3 | 2.874 | 0.165 | -489.110 | 387.84 | .2839 | .7977 | 72.49 | 87.04 | 80.80 |
| | Intercept | -1.438 | 0.102 | | | | | | | |
| 4 | m4 | 3.233 | 0.214 | -500.986 | 364.09 | .2665 | .7625 | 57.58 | 94.92 | 78.90 |
| | Intercept | -1.091 | 0.086 | | | | | | | |
| 5 | m5 | 3.183 | 0.194 | -484.494 | 397.07 | .2907 | .8406 | 63.87 | 93.17 | 80.60 |
| | Intercept | -1.233 | 0.091 | | | | | | | |
| 6 | m1 | 110.282 | -- | 0.000 | 1366.06 | 1.000 | 1.000 | 100.00 | 100.00 | 100.00 |

| | | |
|---|---|---|
| m2 | 155.709 | -- |
| m3 | 176.625 | -- |
| m4 | 155.877 | -- |
| m5 | 165.393 | -- |
| Intercept | -239.993 | -- |

*Note.* The dependent variable in all 6 logistic regression models is Component 1. $N = 1,000$.

# Application 2: National Institute of Mental Health Schizophrenia Collaborative Study

## Data

A second application of our systematic approach was on data from a well-known data set from a psychiatric clinical trial. Hedeker and Gibbons (27) analyzed data on changes in severity of illness in the National Institute of Mental Health Schizophrenia Collaborative Study to illustrate the influence of missing data with a pattern-mixture approach. The target variable was Item 79, *Severity of Illness*, on the Inpatient Multidimensional Psychiatric Scale (IMPS79). Responses were on a scale of 1 (*normal, not at all ill*) to 7 (*among the most extremely ill*), which were treated in the analysis as continuous variables. Participants ($N = 437$) were grouped by placebo vs. drug; the main measurements occurred at Weeks 0, 1, 3, and 6.

To fully describe the missingness of this application example, we report on the missingness by variable, by sample size, and by missing patterns. We analyzed the missingness on IMPS79_0, IMPS79_1, IMPS79_3, and IMPS79_6 ("H&G data set" hereafter), which have 0.7-23.3% missing data. Of the total 437 cases, 73.5% have no missing data, and 26.5% have at least one pattern of missing data. Given 4 variables, the maximum number of missing patterns in this subset of data would be 15 (i.e., $2^4 - 1$). However, in the H&G data set 8 missing data patterns were represented (see Table 2 for missing patterns and frequencies).

The Hedeker and Gibbons application has the advantage of having a sample size and a percentage of missing data that are quite typical for longitudinal psychological and educational research (23). This study design is also typical for the kinds of randomized controlled trials studied in health sciences and psychology. Additionally, because it was studied in such careful detail by Hedeker and Gibbons, it provides an excellent reference for evaluating the solution obtained by our systematic approach.

Hedeker and Gibbons' own approach reduced the missing patterns to one -- missing at Week 6 or not -- which was found to be non-ignorable due to the significant interaction between dropout and assignment to drug treatment condition on the changes in IMPS79 scores over time ($b$ = -.635, $SE$ = .196, $p$ = .002; p. 71). Specifically, for the completers (i.e., observed through week 6), across all four measurements both patients on drug and on placebo experienced decreases in IMPS79 scores, and the decrease was greater for the completers on drug than for those on placebo. For the dropouts (i.e., missing at week 6), the patients on drug experienced significant decrease in IMPS79 scores, whereas those on placebo did not experience much decrease.

## Procedure

We sought to reduce the missing data patterns of the H&G data set using our systematic approach. First, we recoded IMPS79_0, IMPS79_1, IMPS79_3, and IMPS79_6 into corresponding missing data indicators, IMPS79.0m, IMPS79.1m, IMPS79.3m, and IMPS79.6m (Missing = 1, Observed = 0), performed principal component analysis on the missing data indicators, used the four methods to determine the number of principal components to extract, and decided the most appropriate number of components to extract from these data (Steps 1-3). Next, we extracted the component, calculated the predicted component scores, and dichotomized the component scores to make a dichotomous variable that indicates data missing or not (Steps 4-5). Finally, to examine how well this main missing pattern captures the overall missingness on IMPS79_0, IMPS79_1, IMPS79_3, and IMPS79_6, we analyzed the changes in IMPS79 scores by drug, our component, and the interaction between drug and component, and compared our findings to those by Hedeker and Gibbons.

## Results

In Study 1, all four determination methods (i.e., EKC, Kaiser, parallel analysis, and profile likelihood) correctly determined the number of components for more than 99% of the simulated data by similar characteristics to the H&G data set -- $N = 250$, 1 principal component, 3 or 5 variables per component (see row 1 of Figures 4-6 and Supplementary Table 1). Among the four methods, parallel analysis and profile likelihood correctly determined the number of components for 100% of the simulated data. Thus, we expected one principal component to be extracted with the H&G data set (i.e., $N = 437$, 4 variables, and 1.5–54.6% of missing observations). We conducted principal component analysis on the four missing data indicators and found that the four criteria unanimously indicated one component (Table 2). This finding supports our expectation based on Study 1, and provides empirical support for Study 1 findings. As presented in Table 6, missing at week 0 did not load significantly on the principal component ($\beta$ = -.130, $p$ > .05), missing at week 1 significantly but weakly loaded on the component ($\beta$ = -.130, $p$ < .05), but missing at weeks 3 and 6 significantly and strongly loaded on the component ($\beta$ = .863 and .838, respectively, $p$ < .001). The high loadings of IMPS79.3m and IMPS79.6m indicate that the principal missing pattern was mainly driven by missing at Weeks 3 and 6. This result is to a large extent consistent with the pattern (i.e., missing at Week 6) found by Hedeker and Gibbons, but suggests that differences may begin to emerge between completer and attritter may be discernible by week 3.

**Table 6. Application 2: H&G data set, steps 2-3: Missing data indicator component loadings.**

| Component Indicator | Component 1 |
| --- | --- |
| IMPS79.0m | -.130 |
| IMPS79.1m | .300 [a] |
| IMPS79.3m | **.863** [c] |
| IMPS79.6m | **.838** [c] |

[a] $p$ < .05, [b] $p$ < .01, [c] $p$ < .001.  Boldface indicates a loading > .550.

We saved the estimated component scores, and recoded the scores into a dichotomous variable, *Component 1* (below or equal to 0 = 0, Not Missing; above 0 = 1, Missing). Instead of carrying out Steps 6-8, which is recommended for situations when the influence of missing data is to be found out with exploratory analyses, we compared our missing data pattern to a known reference -- the pattern found by Hedeker and Gibbons. Specifically, we modeled the changes in IMPS79 scores from week 0 to week 6 using mixed models with three covariates: (a) Drug (on drug vs. on placebo), (b) Component 1 (missing vs. observed at Weeks 3-6), and (c) the Drug × Component1 interaction. Results indicated that both the main effect of Drug and the Drug × Component1 interaction were significant on changes in IMPS79 scores ($b$ = -.521, $SE$ = .127, $p$ <

.001; $b$ = -.514, $SE$ = .225, $p$ = .022, respectively), whereas the main effect of Component 1 on changes in IMPS79 scores was nonsignificant ($b$ = -.113, $SE$ = .188, $p$ = .547). These findings were consistent with those by Hedeker and Gibbons. Our component 1 (i.e., missing at weeks 3-6) was associated with changes in IMPS79 scores, in the form of an interaction with Drug. As shown in the prototypical trajectories (Figure 11), for the completers, participants on placebo and on drug started very similarly at week 0 but those on drug had a significantly greater decrease in IMPS79 scores. For the dropouts (missing since week 3), despite similar starting points, their difference in decreases was predicted to be much larger than that with the completer pair, favoring the dropouts among individuals assigned to the drug treatment. Our findings highlight the crucial role of missingness at Weeks 3-6 in the estimated decrease of illness severity, as the missingness interacted with the treatment condition, creating differential effects of the drug for those who completed all waves vs. those who were missing since week 3. Our missing data pattern -- although obtained by a very different approach -- also suggests the data are MNAR and contributes to the differential effects of assignment to the intervention group, which is quite consistent with findings by Hederker and Gibbons.

**Fig. 11. Application 2: H&G data set, Steps 6-8: Examining the dichotomized components regarding its influence on changes in IMPS79 scores over time -- completers on placebo vs. dropouts on placebo vs. completers on drug vs. dropouts on drug.**

## Application 3: Depressive Symptoms in the Health and Retirement Study

### Data

The final application considered data on depressive symptoms in the Health and Retirement Study (HRS; (28)). We used the interview data on depressive symptoms from 10 waves (1992 through 2008) of the Health and Retirement Study (HRS). The HRS is a large, nationally representative multi-wave survey containing detailed information on a wide range of topics. Baseline data for the HRS were first collected in 1992 through face-to-face interviews conducted with respondents aged 51-61 (birth cohort 1931-1941) and their spouses. The original sample consisted of 7,600 households and more than 12,600 persons and was based on a multi-stage area probability design oversampling Hispanics, African Americans, and Florida residents. The response rate was over 80% (see  for further details). Follow-ups by phone are performed biannually, with proxy exit interviews for deceased individuals. Beginning in 1998, additional cohorts were added to the sample, drawn from Aging and Health Dynamics in the Oldest Old (AHEAD, born 1923 or before), Children of the Depression Age (CODA, 1924-1930) and War Babies (WB, 1942-1947). The total sample size in the HRS is 31,169.

Our primary variables were depressive symptoms as measured by the Center for Epidemiological Studies Depression (CES-D) scale (29). The HRS uses a subset of 8-items from the original 20-item version of the Center for Epidemiological Studies Depression (CES-D) scale. In terms of the structure of the CES-D items assessed in the HRS, there is representation from the three most important of four CES-D subscales. Specifically, there are three items (depressed, lonely, sad) from the depressed affect (DA) subscale, two items (happy, enjoyed life) from the positive affect (PA) subscale, and three items (effort, sleep, get going) from the somatic complaint (SO) subscale. (Interpersonal symptoms are not represented.) The resulting variables examined in the present study were subscale scores on depressed affect (DA), positive affect (PA), and somatic complaint (SO) for each wave, which makes a total of 30 variables over 10 waves ("CES-D data set" hereafter).

To fully describe the missingness of this application data set, we report on the missingness by variable, by sample size, and by missing pattern. There are 38.1-67.2% missing data across the 30 variables. Of the 31,169 cases in the HRS study, 3,868 (12.4%) cases have missing data on all 30 variables, which results in an analytic sample size of 27,301 for the CES-D data set. Of the analytic sample, 18.2% have no missing data, and 81.8% have missing data for at least one variable. Given the 30 variables, the maximum number of missing patterns would be 1,073,741,823 (i.e., $2^{30} - 1$). We observed 898 different missing data patterns in this subset of data, which is far from the maximum but still an unwieldy number of missing patterns in this data set (see Table 2 for missing patterns with the highest and lowest frequencies).

The CES-D data set provides the opportunity to evaluate our systematic approach with data on a large number of variables derived from a large number of waves with a large sample of 27,301 participants. A further advantage of considering the data from the HRS is that it contains known study design features and participant status in the HRS that account for some of the missing values. For example: 1) cohorts were added into the study in various waves; 2) vital status is known for this sample of older adults; 3) physical or cognitive impairment precluded completion of various self-report measures for some individuals in some waves. This information is housed in a separate study-wide tracking database -- the HRS Tracker file, which contains demographic information, interview status, and if, when and how an interview was conducted in each wave. Therefore, we could compare the observed missing data patterns in the CES-D data set (i.e., whether data were actually observed) with the participant status information in the Tracker file (i.e., whether participants were expected to provide data in each wave). Our systematic approach extracted main missing data patterns independently of information in the Tracker file, which was used only after the fact in order to validate and interpret our extracted missing data patterns.

## Procedure

We sought to reduce 898 missing data patterns with our systematic approach. First, we recoded the 30 CES-D variables into 30 corresponding missing data indicators, DA1m to SO10m (Missing = 1, Observed = 0), performed a principal component analysis on the missing data indicators, used the four methods to determine the number of principal components to extract, and decide the most appropriate number of components to extract from these data (Steps 1-3). Next, we extracted the components, calculated the predicted component scores, and dichotomized the principal component scores to make a binary variable indicating whether variables were missing or observed (Steps 4-5). As such, we have reduced the 898 missing data patterns to 4 patterns (see Results for details).

To examine how well the missing data patterns captured the overall missingness in the 30 CES-D variables (Steps 6-8), we conducted logistic regressions of each dichotomous component on 10 index variables (i.e., *index01* to *index10*) that informs whether a participant was expected to provide data on these variables at each wave. We created these 10 variables based on variables in the HRS Tracker file, including participant's vital status (i.e., *alive*), whether a proxy provided interview data for the participant (i.e., *proxy,* which indicates that self-reported CES-D data should *not* be collected), whether interview data was obtained in a given wave (i.e., *iwwave*), and the year when first interviewed (i.e., *firstiw*). Each of the 10 composite variables shows whether or not a participant's CES-D data *should* be missing at a measurement occasion for design reasons (Should be Missing = 1, Should be Observed = 0). For instance, the variable, *index03*, represents whether or not a participant's CES-D data should be missing in 1996 (Wave 3 of the HRS): if a participant was interviewed before or in 1996 for the first time, and alive in 1996, her interview data were self-reported within this wave, and her interview data were not provided by a proxy, then this participant's CES-D data should not be missing in 1996, and her *index03* = 0 (i.e., Should be Observed); if any of the above-mentioned conditions was not met, then her *index03* = 1 (i.e., Should be Missing; see Table 6 for frequencies of *index01* to *index10*).

**Table 6. Application 3: CES-D data set, Steps 6-8: External indices of HRS participation status.**

| Variable | Year | Status | $N$ | Percentage % |
|---|---|---|---|---|
| index01 | 1992 | Should be observed | 12,004 | 38.5 |
|  |  | Should be missing | 19,165 | 61.5 |
| index02 | 1994 | Should be observed | 10,691 | 34.3 |
|  |  | Should be missing | 20,478 | 65.7 |
| index03 | 1996 | Should be observed | 10,225 | 32.8 |
|  |  | Should be missing | 20,944 | 67.2 |
| index04 | 1998 | Should be observed | 19,341 | 62.1 |
|  |  | Should be missing | 11,828 | 37.9 |
| index05 | 2000 | Should be observed | 17,517 | 56.2 |
|  |  | Should be missing | 13,652 | 43.8 |
| index06 | 2002 | Should be observed | 16,130 | 51.8 |
|  |  | Should be missing | 15,039 | 48.2 |
| index07 | 2004 | Should be observed | 18,327 | 58.8 |
| | | Should be missing | 12,842 | 41.2 |
| index08 | 2006 | Should be observed | 17,209 | 55.2 |
| | | Should be missing | 13,960 | 44.8 |
| index09 | 2008 | Should be observed | 16,077 | 51.6 |
| | | Should be missing | 15,092 | 48.4 |
| index10 | 2010 | Should be observed | 14,164 | 45.4 |
| | | Should be missing | 17,005 | 54.6 |

*Note.* Index01-Index10 were created using a number of variables in the Tracker Database of the HRS ($N = 31,169$) independent of the CES-D data set. These variables include vital status, whether interviewed within wave, proxy status, and year of first interview. Should be observed = 0, Should be missing = 1.

## Results

In Study 1, three determination methods -- EKC, Kaiser, and parallel analysis -- correctly determined the number of components for 100% of the simulated data with similar characteristics to the CES-D data set -- $N = 1000$, 3 or 5 principal components, 5 or 10 variables per component, and 50% missing data (see rows 2-3 of Figures 7-9 and the Supplementary Table). In Study 1, the largest sample size we simulated was $N = 1,000$, so we compared the CES-D data set with these simulated data. Even though 1,000 is not "similar" to 27,301, the larger CES-D sample size is of higher statistical power, which results in a higher correct rate. The performance with data sets of $N = 1,000$ was extremely close to 100% correct, therefore, the performance with the CES-D data set would be better but not worse than close 100% correct, given similarities in other aspects, i.e., the number of principal components, the number of items per component, and percentage of missing data. Thus, we expected 3, 4, or 5 components to be extracted with the CES-D data set (i.e., $N = 27{,}301$, a total of 30 variables, and 38.1-67.2% missing observations) based on the more accurate EKC, Kaiser, or parallel analysis methods. We conducted principal component analysis on the 30 missing data indicators, and found that three methods unanimously indicated 4 components, which is different from what profile likelihood suggested (Table 3). This finding confirms our hypothesis based on Study 1, and provides empirical support for the Study 1 findings. As presented in Table 7, the missing indicators loaded on corresponding components with a standardized loading higher than .560.

**Table 7. Application 3: CES-D data set, steps 2-3: Missing data indicator component loadings.**

| Component Indicator | Component 1 | Component 2 | Component 3 | Component 4 |
|---|---|---|---|---|
| DA1m | -.039 | -.087 | -.008 | **.985**[c] |
| SO1m | -.039 | -.087 | -.008 | **.984**[c] |
| PA1m | -.039 | -.087 | -.008 | **.985**[c] |
| DA2m | .003 | .007 | .006 | **.958**[c] |
| SO2m | .003 | .007 | .006 | **.958**[c] |
| PA2m | .003 | .006 | .006 | **.958**[c] |
| DA3m | .048 | .099 | -.014 | **.904**[c] |
| SO3m | .047 | .099 | -.013 | **.903**[c] |
| PA3m | .048 | .099 | -.014 | **.904**[c] |
| DA4m | .055 | **.979**[c] | -.160[b] | .014 |
| SO4m | .055 | **.979** [c] | -.160 [b] | .014 |
| PA4m | .055 | **.979** [c] | -.161 [b] | .014 |
| DA5m | .000 | **.880** [c] | .121 [b] | .012 |
| SO5m | .000 | **.880** [c] | .121 [b] | .013 |
| PA5m | .000 | **.880** [c] | .121 [b] | .012 |
| DA6m | -.071 | **.641** [c] | *.472* [c] | .064 |
| SO6m | -.069 | **.641** [c] | *.471* [c] | .063 |
| PA6m | -.071 | **.641** [c] | *.471* [c] | .065 |
| DA7m | .140 [b] | .052 | **.840** [c] | .017 |
| SO7m | .142 [b] | .052 | **.839** [c] | .016 |
| PA7m | .140 [b] | .051 | **.840** [c] | .015 |
| DA8m | *.474* [c] | .006 | **.566** [c] | .023 |
| SO8m | *.476* [c] | .005 | **.563** [c] | .024 |
| PA8m | *.474* [c] | .006 | **.565** [c] | .023 |
| DA9m | **.788** [c] | .013 | .228 [b] | .012 |
| SO9m | **.789** [c] | .013 | .226 [b] | .013 |
| PA9m | **.788** [c] | .013 | .227 [b] | .012 |
| DA10m | **.992** [c] | .022 | -.076 | .011 |
| SO10m | **.991** [c] | .022 | -.076 | .012 |
| PA10m | **.991** [c] | .022 | -.076 | .012 |

[a] $p < .05$, [b] $p < .01$, [c] $p < .001$. Boldface indicates a loading > .550. Italic indicates .450 < a cross-loading < .550.

We saved the estimated component scores, and recoded the scores into 4 dichotomous variables, *Components 1* to *4* (below or equal to 0 = 0, Not Missing; above 0 = 1, Missing). Component 1 represented missing from waves 9-10 with cross-loadings capturing some missingness at wave 8. Component 2 represented missing from waves 4-6. Component 3 represented missing from waves 7-8 with cross-loadings capturing some missingness at wave 6. Component 4 represented missing from waves 1-3. As such, we have reduced 898 missing data patterns to 4 main patterns for the CES-D data set, each of which is structured around missing data across contiguous waves of data collection.

We conducted four logistic regression models of components 1-4, each on all 10 index variables. All four models converged successfully and provided better fit to the data than the corresponding null models with only the intercept (Table 7). The models all had high sensitivity, specificity, and an overall rate of correct classification close to 1. As presented in Table 8, the regression coefficients were consistent with the principal component analysis results in suggesting that Component 1 is linked with those who should be missing at waves 9-10, Component 2 associated with those who should be missing at waves 4-6, Component 3 associated with those who should be missing at waves 7-8, and Component 4 linked with those who should be missing at waves 1-3. For example, a participant had index01 = 1, index02 = 1, and index03 = 1, which means she should not be providing data at waves 1-3. Our approach showed that she had Component 4 = 1 and all other components = 0, correctly identifying her CES-D scores at waves 1-3 were missing. That is, according to study design information from the HRS Tracker file, this participant was not enrolled in the HRS until wave 4. Our systematic approach, although did not utilize any study design information to obtain the four main missing patterns (i.e., the 4 components), captured the 4 main missing patterns that accurately reflect the missingness caused by the design of the HRS.

**Table 8. Application 3: CES-D data set, Steps 6-8: Logistic regressions of each component on the 10 index variables: Model fit indices and percentage of correction prediction.**

| Dependent Variable | Log Likelihood | LR chi$^2$ (*df*) | Pseudo R$^2$ | Area under ROC Curve | Sensitivity (%) | Specificity (%) | Correctly Classified (%) |
|---|---|---|---|---|---|---|---|
| Component 1 | -512.105 | 42068.210 (10) | .976 | .999 | 99.7 | 99.5 | 99.6 |
| Component 2 | -353.967 | 42134.440 (10) | .984 | 1.000 | 99.8 | 99.9 | 99.8 |
| Component 3 | -602.613 | 41316.560 (10) | .972 | .999 | 99.5 | 99.9 | 99.7 |
| Component 4 | -241.778 | 40214.770 (10) | .988 | 1.000 | 99.8 | 99.4 | 99.6 |

**Table 9. Application 3: CES-D data set, Steps 6-8: Logistic regressions of each component on the 10 index variables: Model coefficients.**

| | Component 1 | | Component 2 | | Component 3 | | Component 4 | |
|---|---|---|---|---|---|---|---|---|
| | Coef. Est. | S.E. | Coef. Est. | S.E. | Coef. Est. | S.E. | Coef. Est. | S.E. |
| Intercept | -11.714[c] | 0.515 | -16.008[c] | 0.629 | -7.194[c] | 0.256 | -16.744[c] | 0.985 |

| index01 | -1.689[c] | 0.412 | -1.744[c] | 0.452 | 0.294 | 0.439 | **5.259**[c] | 0.455 |
| index02 | 0.259 | 0.408 | -0.402 | 0.472 | -0.394 | 0.457 | **11.964**[c] | 0.724 |
| index03 | 0.972[b] | 0.362 | 2.226[c] | 0.340 | -1.110[b] | 0.347 | **11.433**[c] | 0.737 |
| index04 | 0.867[b] | 0.290 | **18.803**[c] | 0.719 | -3.904[c] | 0.277 | -0.586 | 0.338 |
| index05 | -1.009[c] | 0.288 | **13.067**[c] | 0.551 | -0.126 | 0.246 | 0.252 | 0.384 |
| index06 | -5.989[c] | 0.298 | **12.731**[c] | 0.541 | **6.070**[c] | 0.260 | -0.579 | 0.401 |
| index07 | -4.187[c] | 0.287 | -3.663[c] | 0.358 | **16.095**[c] | 0.521 | 0.004 | 0.441 |
| index08 | **7.023**[c] | 0.349 | -1.530[c] | 0.350 | **8.352**[c] | 0.291 | 0.648 | 0.433 |
| index09 | **8.112**[c] | 0.374 | -0.153 | 0.362 | 1.216[c] | 0.260 | -0.058 | 0.421 |
| index10 | **16.400**[c] | 0.611 | 0.726[a] | 0.317 | -5.667[c] | 0.294 | -0.474 | 0.390 |

*Note.* [a] $p < .05$, [b] $p < .01$, [c] $p < .001$. Boldface indicates the 3 most influential predictors for each component.

## Discussion

Growing accessibility of methods for analysis of missing data mechanisms has not been matched by consensus on methods for systematically evaluating the assumptions on which successful application of these methods rest. In biostatistics, where a single outcome is typically of interest, the missing data mechanism is relatively straightforward to explore. In psychological, social, and educational research, data are often multivariate and longitudinal, which requires a reasoned approach to data reduction prior to examination of missing data mechanisms, since the number of potential missing data patterns grows exponentially with the number of variables in the analysis.

We evaluate principal component analysis with a number of commonly applied methods for determining the number of components to retain, using simulated and real data examples representing the broad spectrum of conditions encountered in psychological and educational research. Under nearly all conditions, our results point to a number of clear and concrete recommendations.

In terms of evaluating performance of various methods for determining the number of components to extract, our results align well with the general consensus in the literature. Further, because we are primarily interested in the number of components rather than the component loadings, our results suggest that these methods work just as well with Pearson's $r$ as with more computationally intensive tetrachoric correlations.

Consistent with the broader literature on component extraction, our results suggest that parallel analysis should be the preferred method under most circumstances. Using simulated data, this method recovers the correct number of components in over 95% of cases under every condition we studied for sample sizes of at least 250. Because this intermediate sample size was selected on the basis of a "typical" study in psychological and educational research, this suggests that the approach we outline here is likely to be effective in a majority of circumstances encountered by psychological and educational researchers.

We also find evidence that this approach is likely to work well in practice, using one simulated data set generated according to an unknown mechanism known to be MNAR, a data set with demonstrated MNAR mechanism that has received careful analysis in the literature, and a large longitudinal population-based study where several design characteristics partially determine the patterns of missing data.

An approach such as the one outlined here is also likely to be useful for public-use data sets that are increasingly distributed with multiple imputations for missing values. Several examples exist of a multiple imputation approach with large scale data sets (30–32). An approach similar to the one we outline here could be very useful when coupled with popular approaches such as multiple imputation using chained equations (33). Whereas some authors have favored a "kitchen sink" approach to imputation, e.g., (3), we would expect that a more limited approach would ultimately be favored in the bias-variance tradeoff (34,35).

Missing data always introduce uncertainty about unobserved values. In the best circumstances, they do so in a similar fashion to the way that analysis of data from a sample introduces uncertainty about parameters in the population, but informative non-response can have considerable consequences for study conclusions if not more fully evaluated. Further, methods for the analysis of MNAR data are still emerging and this problem has been shown theoretically to be intractable based only on the observed data (36). For this reason, recommendations for analyses with missing data increasingly revolve around sensitivity analyses. This shifts the emphasis toward methods that reduce, or at least better characterize, the extent of potential bias for key predictors of interest. Of particular interest is the extent to which study conclusions depend on the assumptions, implicit or explicit, being made about the nature of unobserved values.

The approach we recommend here is strongly consistent with a sensitivity analysis framework in two ways. First, examination of associations with a smaller number of missing data patterns (i.e., the principal components) allows each of them to be considered in greater depth. We have outlined a minimal set of specific and systematic recommendations that can be used to guide this process for a given number of missing data components.  Second, we also advocate consideration of different numbers of missing data components (e.g., one more and one fewer than extracted on the basis of parallel analysis). In many cases, results agree very closely across these conditions. In circumstances where results diverge, this is important information that can be used to guide sensitivity analyses and to appropriately qualify results.

## Limitations

We have considered a broad subset of conditions that reflect the range of most psychological research. However, researchers might want to conduct similar simulations that more closely align with their own data. In real data applications, we do not know the ground truth, and so cannot truly evaluate performance with real data. However, our results with real data are both broadly interpretable and consistent with previous analyses that are available for two of the data sets we considered here, and the simulation model we used here can easily be applied to any specific combinations of factors of interest.

Another important consideration is that study results, particularly adjustments for factors associated with the missing data mechanism, are only going to be as good as the quality of ancillary data available. This might suggest that researchers devote greater attention to collecting additional follow-up data from participants as possible, something that has long been recommended practice in the field of survey research (37).

# Future Directions

This study presents one validated approach to promote consideration of missing data mechanisms in multivariate and/or longitudinal contexts, but considerable opportunities remain with regard to better aligning missing data theory and practice, particularly for highly multivariate. For example, our missing data mechanisms had known structure for the simulated data, but not for the real data applications. Future research should examine the performance of these procedures using real data with deliberately masked observations.

Similarly, in many areas of research, such as analysis of clinical trials data, there is appropriate concern on missing data methods that can be "gamed" or manipulated by the data analyst. This can sometimes lead to widespread adoption of some less favorable but objective treatments of missing data, such as last observation carried forward. Although this is one of our goals in presenting this approach in as systematic and replicable a manner as possible, and situating it explicitly within a sensitivity analysis framework, future research should carefully evaluate where missing data protocols need to be modified in order to further remove or reduce subjectivity in the application of missing data methods.

In summary, we presented a standardized procedure for exploring missing data mechanisms with multivariate and/or longitudinal data. We evaluated several commonly used criteria for selecting the number of principal components using dichotomous missing data indicators as one tool for guiding investigation of potential missing data mechanisms and sensitivity analyses when considering the assumptions about missing data for study conclusions.

Acknowledgements

Research reported in this publication was supported by the National Cancer Institute

(https://www.cancer.gov/) of the National Institutes of Health (https://www.nih.gov/) under

Award Number R01CA194178. The content is solely the responsibility of the authors and does not necessarily represent the official views of the National Institutes of Health. The funding body played no role in the design of the study and collection, analysis, and interpretation of data and in writing the manuscript.

Dai is partially supported by "Inference-making and reasoning: refinement of an assessment for

use in gateway biology courses" funded by the Institute of Education Sciences, U.S. Dept. of

Education: under Award #R305A160335. The opinions expressed are those of the authors and do not represent views of the Institute or the U.S. Department of Education.

The HRS (Health and Retirement Study) is sponsored by the National Institute on Aging (grant

number NIA U01AG009740) and is conducted by the University of Michigan. The opinions and conclusions expressed are solely those of the author and do not represent the opinions or policy of the National Institutes of Health or the HRS.

# Proportions of Reps. with a Correct # of Components Determined
## $N$ = 100, 10% Missing, performing PCA with Pearson $r$

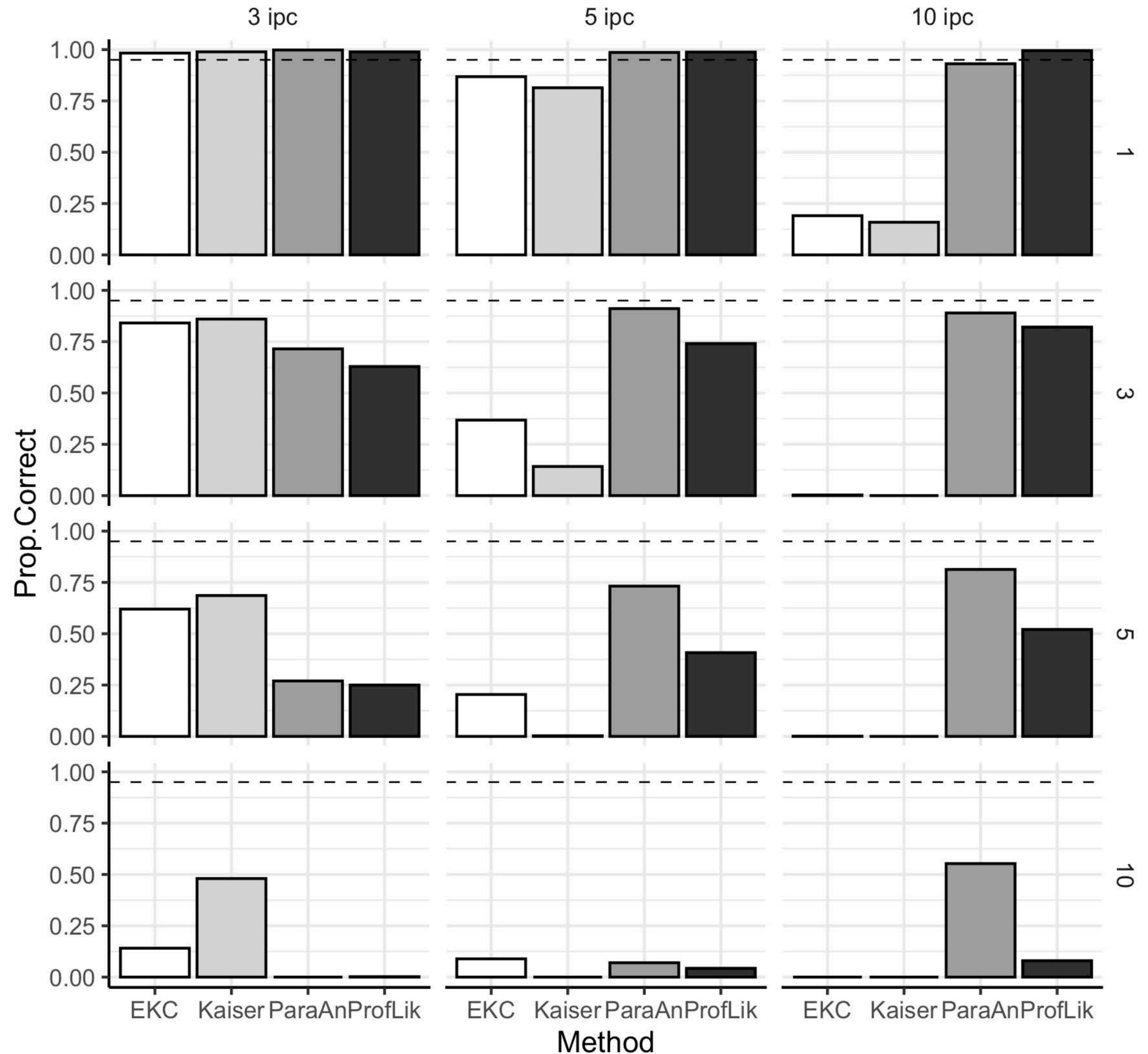

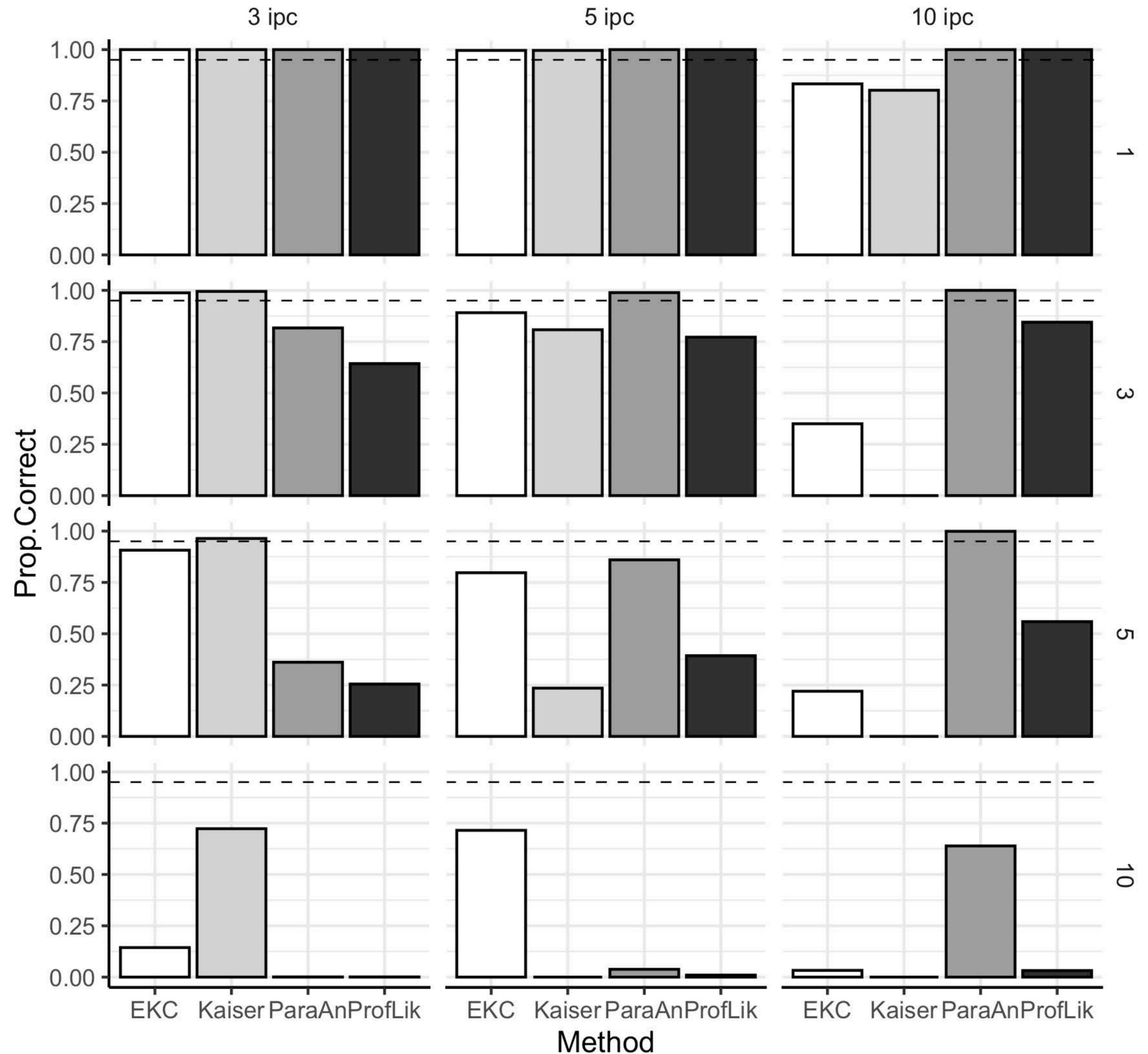

Proportions of Reps. with a Correct # of Components Determined
$N$ = 100, 25% Missing, performing PCA with Pearson $r$

# Proportions of Reps. with a Correct # of Components Determined
## $N = 100$, 50% Missing, performing PCA with Pearson $r$

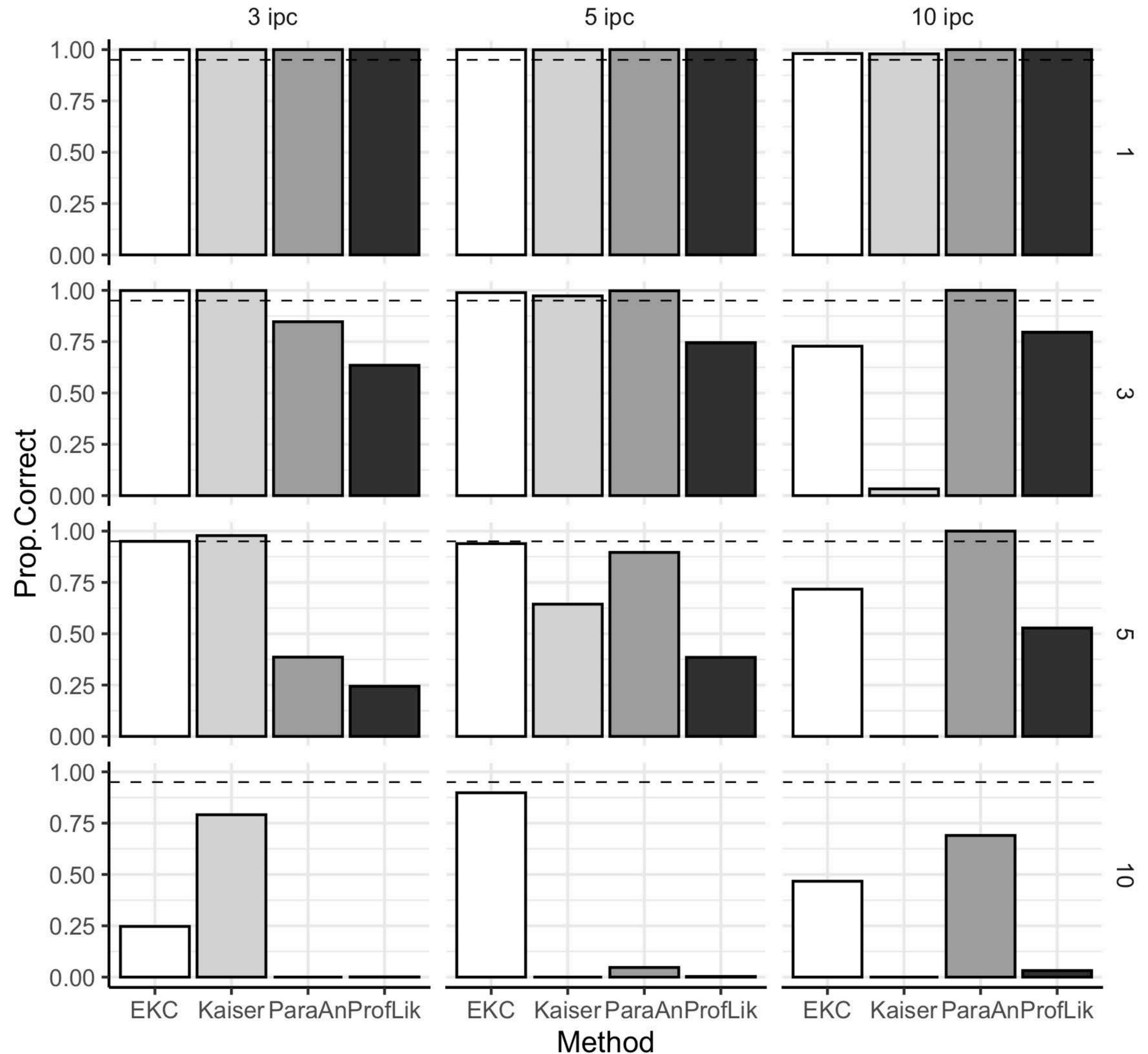

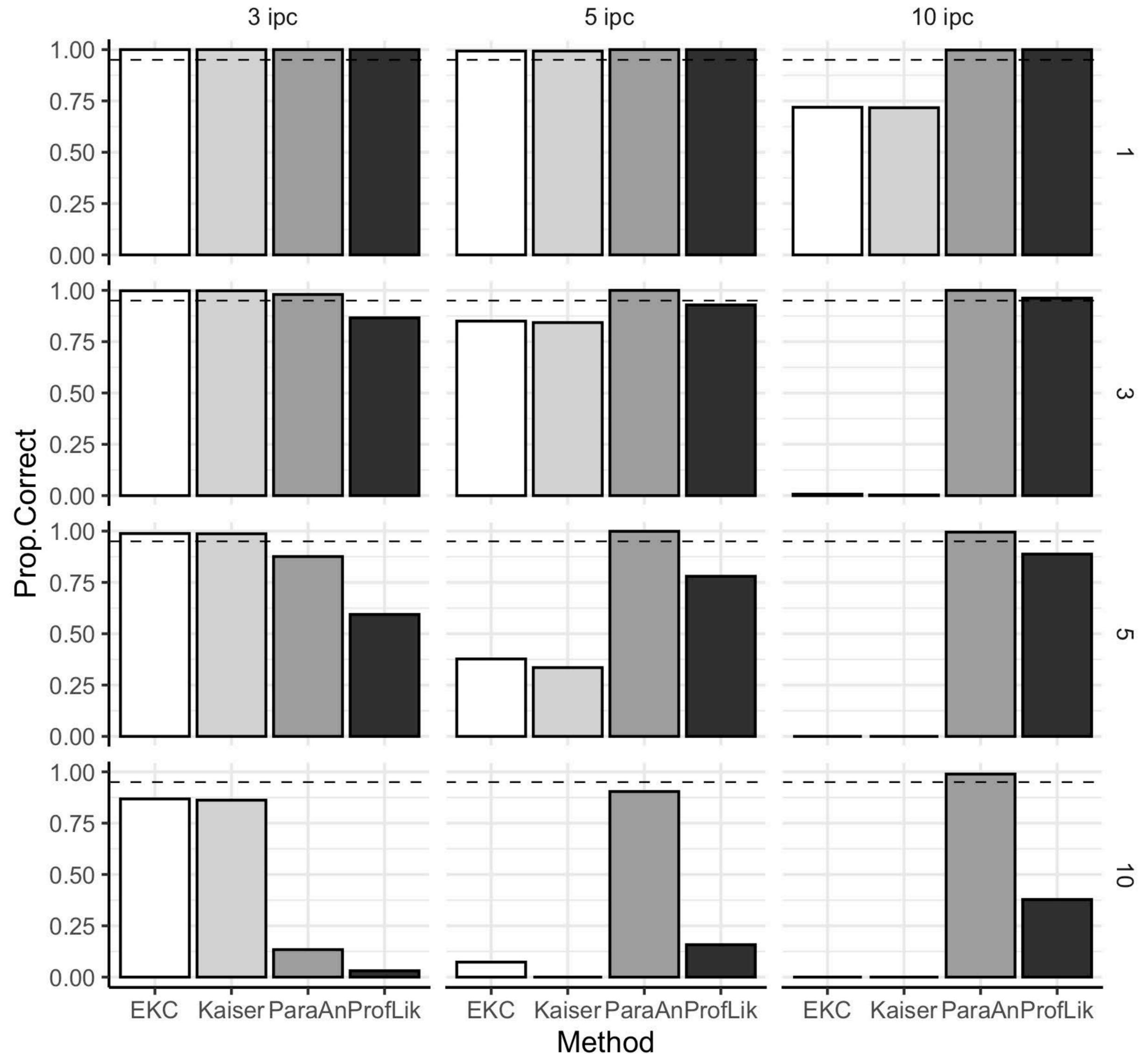

Proportions of Reps. with a Correct # of Components Determined
$N$ = 250, 10% Missing, performing PCA with Pearson $r$

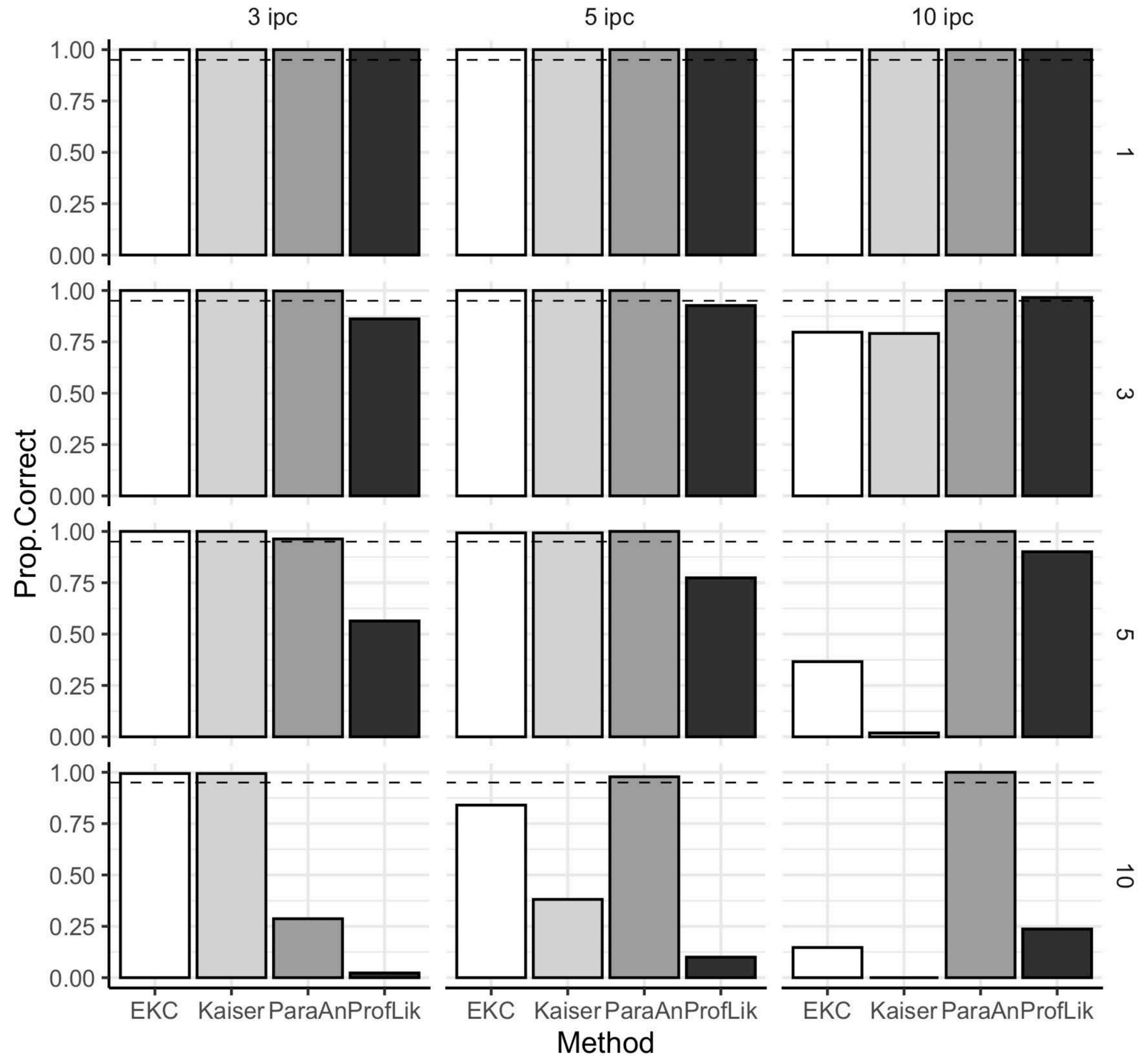

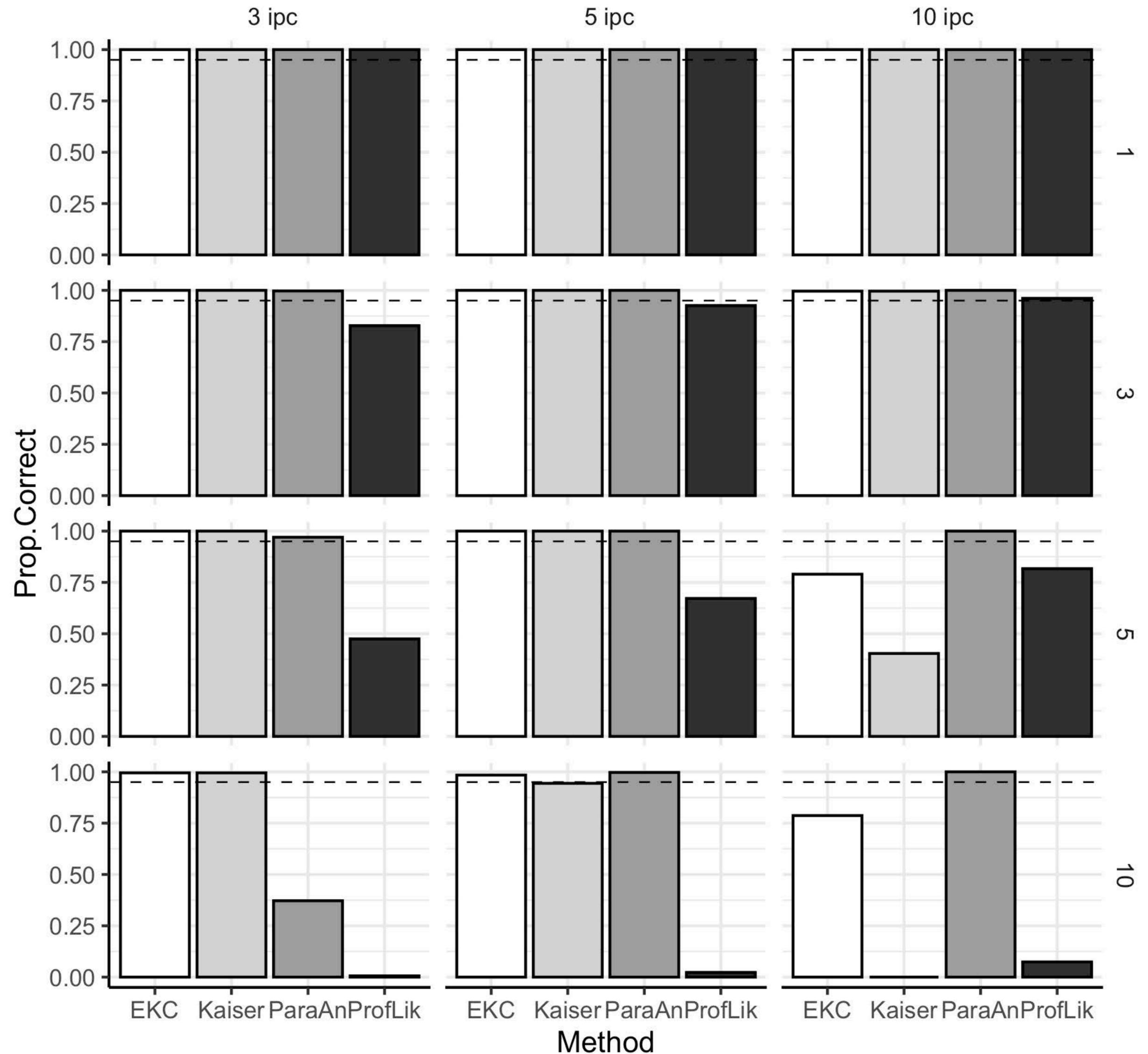

Proportions of Reps. with a Correct # of Components Determined
$N$ = 250, 50% Missing, performing PCA with Pearson $r$

# Proportions of Reps. with a Correct # of Components Determined
## $N = 1000$, 10% Missing, performing PCA with Pearson $r$

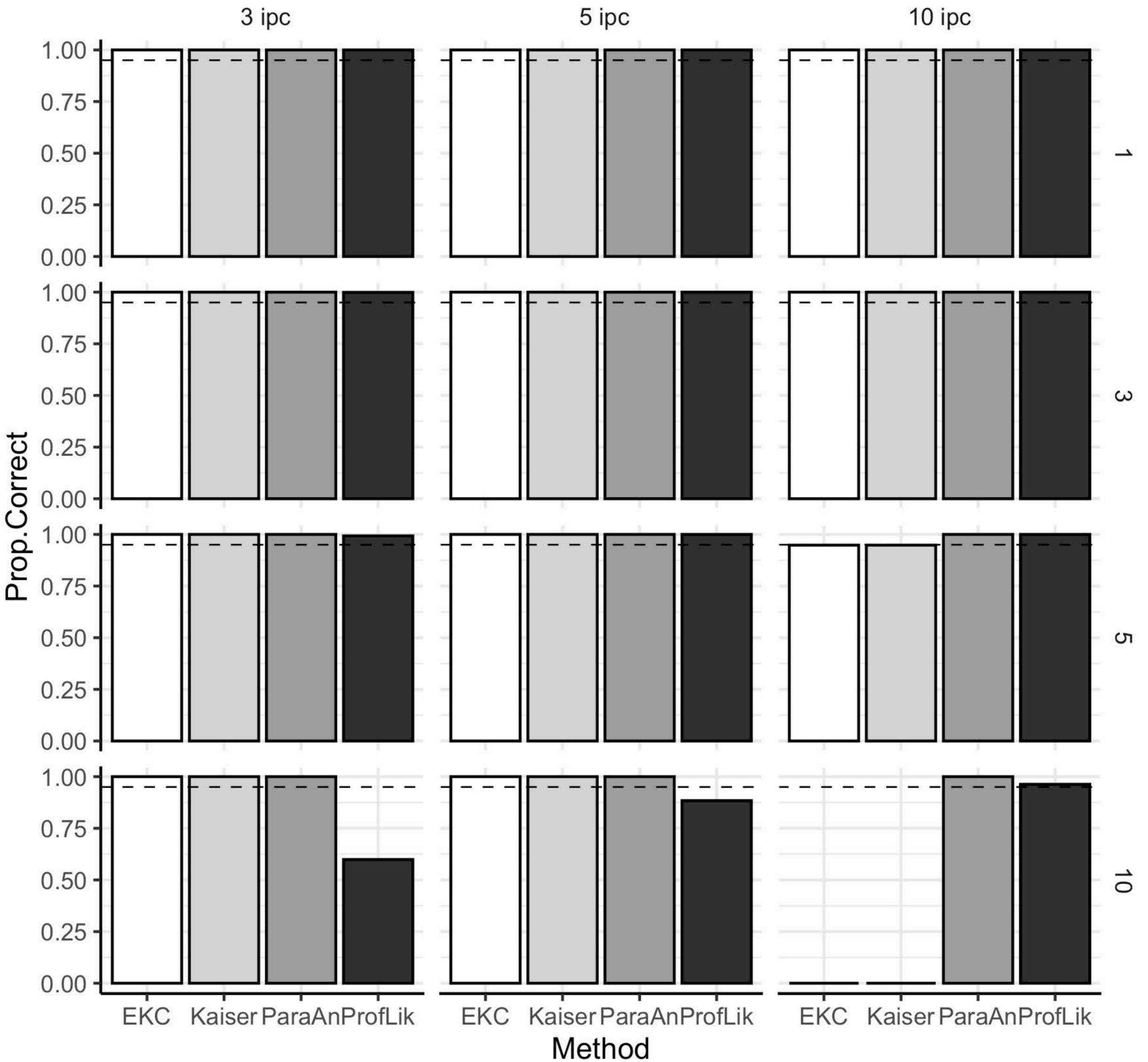

# Proportions of Reps. with a Correct # of Components Determined
## $N$ = 1000, 25% Missing, performing PCA with Pearson $r$

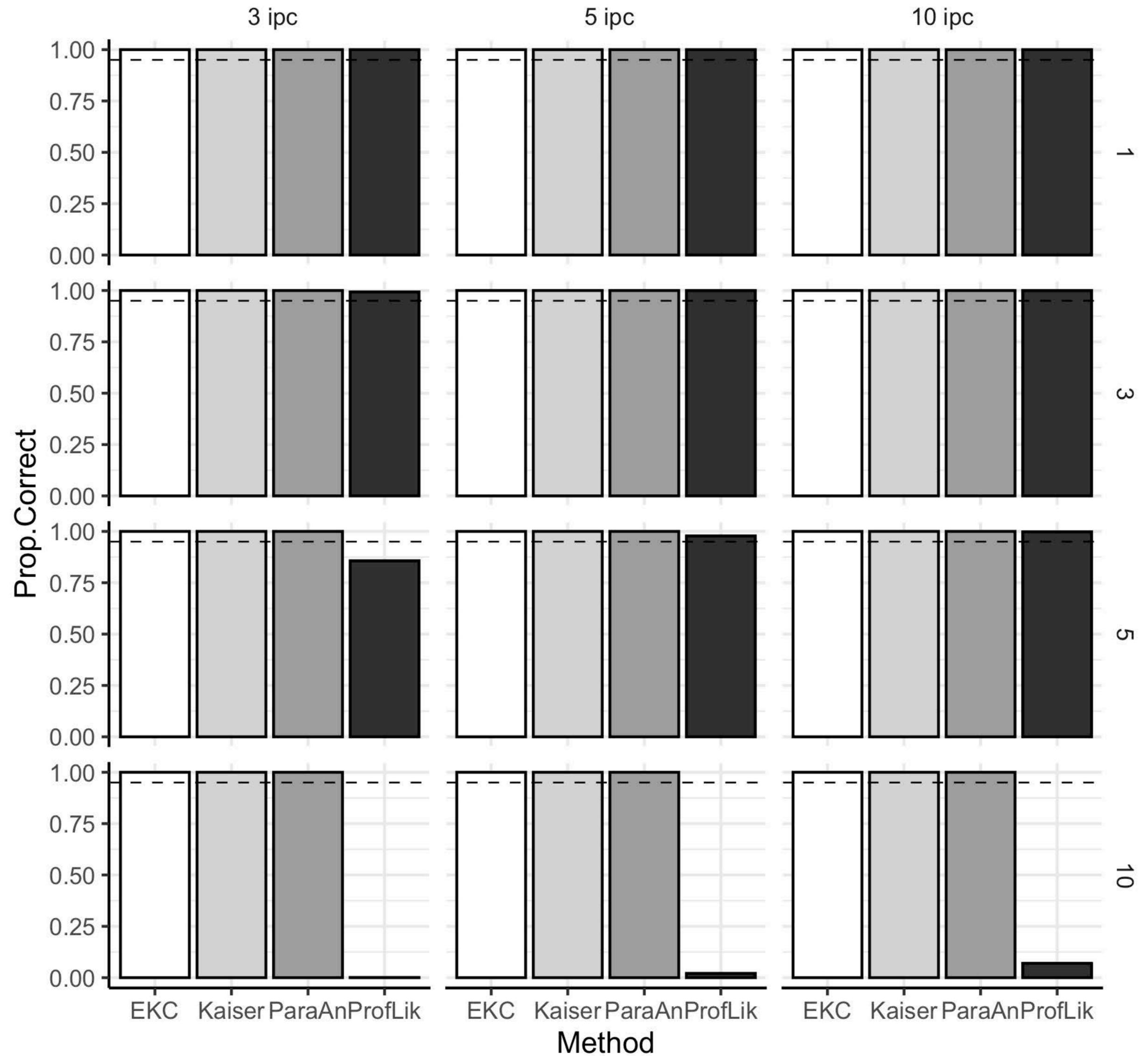

# Proportions of Reps. with a Correct # of Components Determined
## $N$ = 1000, 50% Missing, performing PCA with Pearson $r$

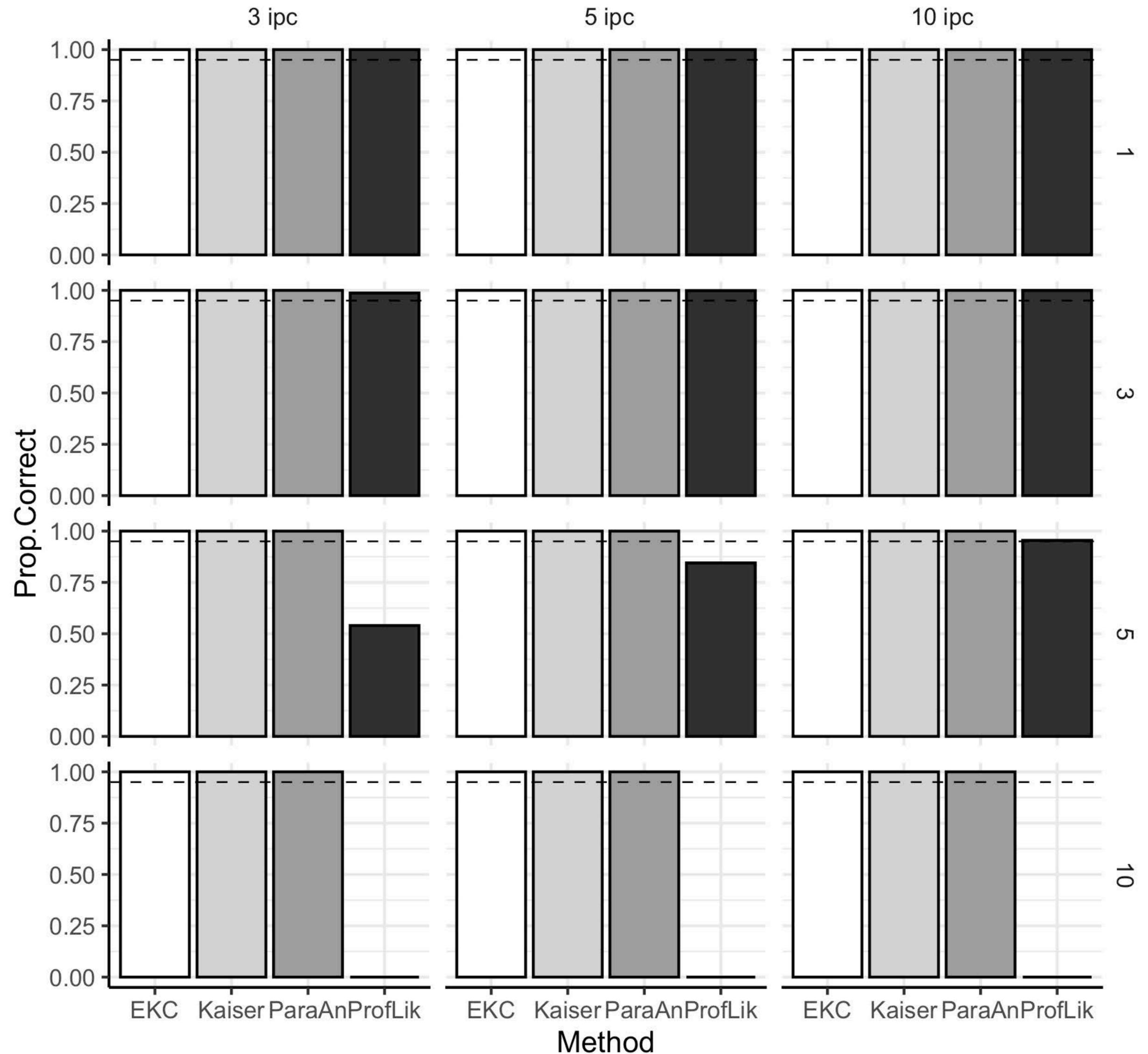

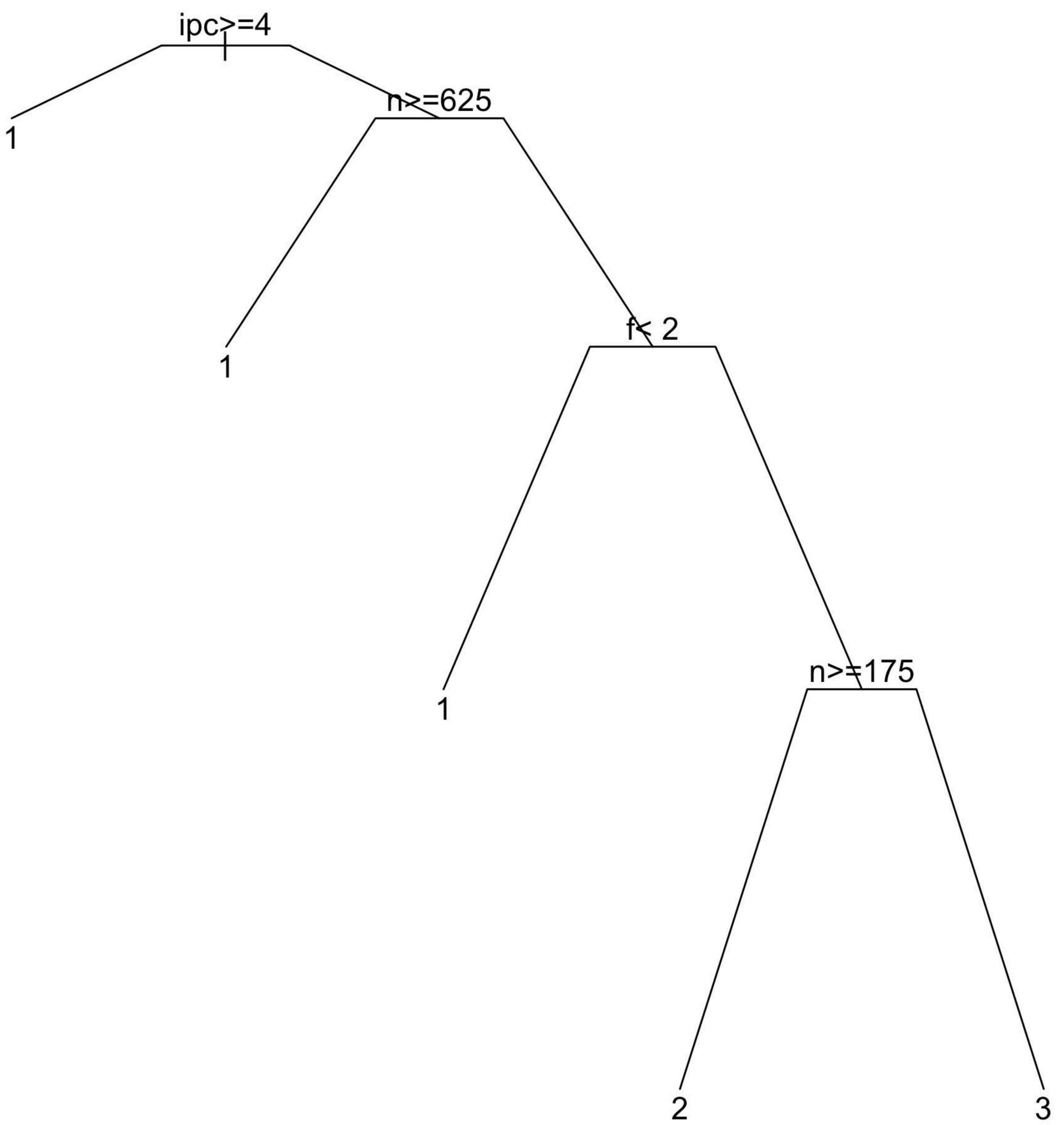

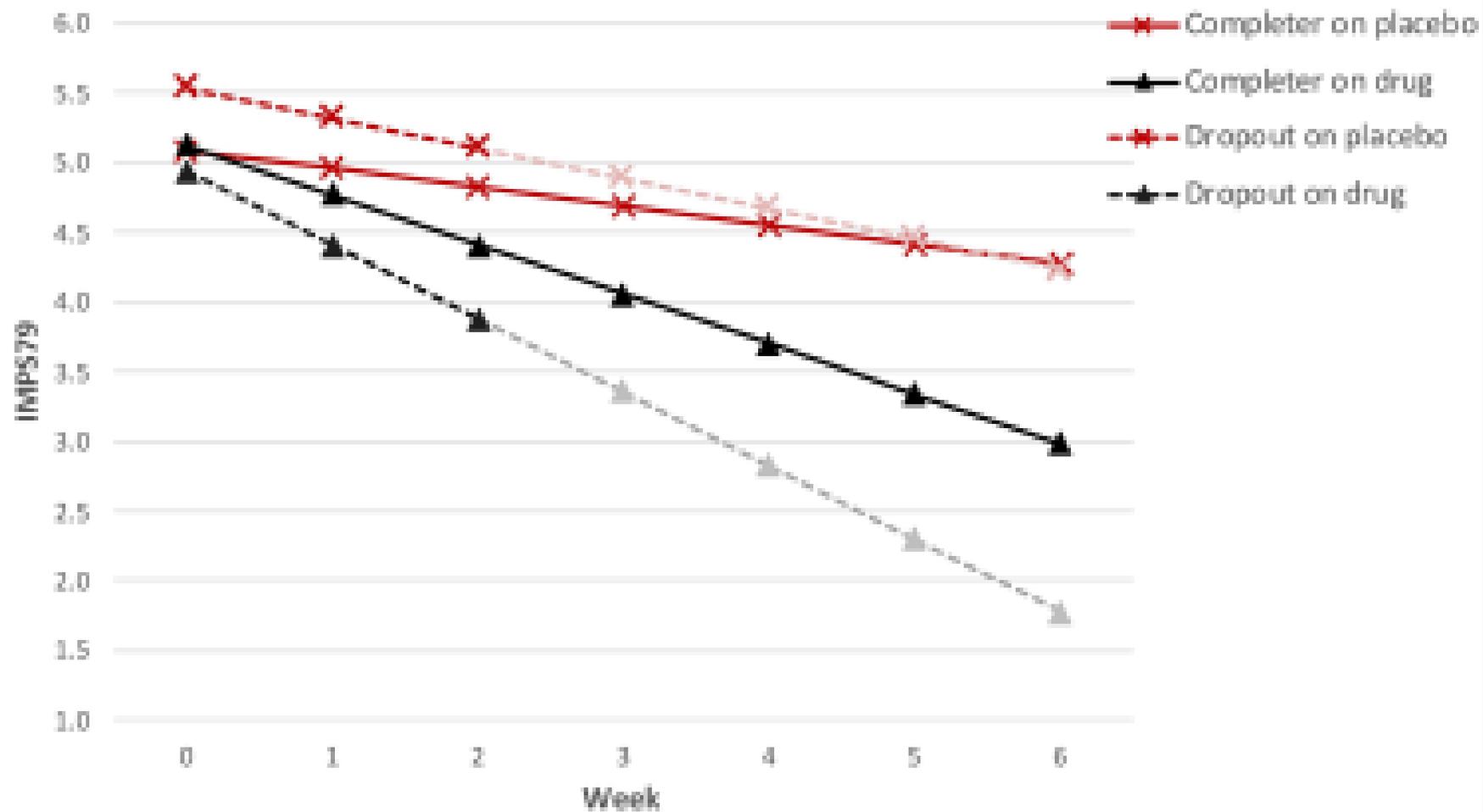